\documentstyle[bo99,epsfig]{article}

\title{The Parkes Multibeam Pulsar Survey: \\
              preliminary results}

\author{
  N. D'Amico$^{1}$, A.G. Lyne$^{2}$, R.N. Manchester$^{3}$, 
  F.M. Camilo$^{2}$, V.M. Kaspi$^{4}$,
  J. Bell$^{3}$, I.H. Stairs$^{2}$, F. Crawford$^{4}$, 
  D. Morris$^{2}$, A. Possenti$^{1}$}

 \affil{1) Osservatorio Astronomico di Bologna, via Ranzani 1, 
    40127 Bologna, \\ 
           and Istituto di Radioastronomia del CNR, via
    Gobetti 101, 40129 Bologna, Italy \\
        2) University of Manchester, Jodrell Bank Observatory, \\ 
         Macclesfield,
       Cheshire, SK11 9DL, UK \\
        3) Australia Telescope National Facility, CSIRO, PO Box 76, \\
           Epping
       NSW, 1710, Australia \\
        4) Massachusetts Institute of Technology, Center for Space Research,
       70 Vassar Street, Cambridge, MA 02139, USA}

\begin{document}
\newdimen\digitwidth    
\setbox0=\hbox{\rm0}
\digitwidth=\wd0
\catcode`!=\active
\def!{\kern\digitwidth}

\maketitle

\begin{abstract}

  A high-frequency survey of the Galactic plane for radio pulsars
  is in progress, using the multibeam receiver on the 64-m Parkes
  radiotelescope.  We describe the survey motivations, the observing 
  plan and the inital results.   The survey is discovering many pulsars, 
  more than 500 so far.  Eight of  the new pulsars are binary,  one with 
  a massive companion. At least eight are young, with characteristic ages 
  of less than 100 kyr. Two of these (Kaspi et al, this Conference) have 
  surface dipole magnetic field strengths greater than any other known 
  radio pulsar.

\keywords{methods: observational; pulsars: general; pulsars: searches.}               
\end{abstract}

\section{Introduction}

Young pulsars are relatively rare objects in the pulsar population because 
they evolve rapidly, so on average their distance is relatively high. They
are usually found at low Galactic latitudes, close to their places 
of birth, where their detection is  limited by the
high background temperature and by the broadening of pulses due to
dispersion and interstellar scattering.  On the other hand, they are
interesting objects for many reasons: they are likely
to be $\gamma$--ray sources; they exibhit rotational {\it glitches},
which are of interest in the understanding of the interior structure
of neutron stars; they are likely to be associated with supernova remnants.

\begin{figure}
\centerline{\psfig{file=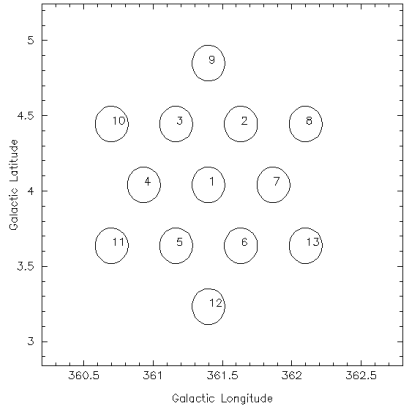, width=7cm}}
\caption[]{Beam pattern of the multibeam receiver at Parkes.}
\end{figure}

High frequency ($\nu$ $\simeq$ 1400 MHz) surveys (Clifton et al, 1992; Johnston et
al 1992) and searches (Kaspi et al, 1996; Manchester, D'Amico \& Tuohy, 1985) 
for young, low latitude distant 
pulsars proved to be successful, because the
contribution of the Galactic synchrotron radiation to the radiotelescope
system temperature is highly reduced, because the effect of dispersion is 
more easlily removed, and because the broadening of pulses due to interstellar
scattering varies with frequency aproximately as $\nu^{-4.4}$.
\begin{table}
\caption{Three 20\,cm pulsar surveys}
\begin{tabular}{llll}
\hline
                               & Jodrell Bank     & Parkes      & Parkes     \\
\hline
Latitude range, $|b|$\dotfill  & $<1^\circ$       & $<4^\circ$  & $<5^\circ$ \\
Longitude range, $l$\dotfill   & $-5^\circ$\dots$100^\circ$
                                 & $-90^\circ$\dots$20^\circ$
                                 & $-100^\circ$\dots$50^\circ$               \\
Center frequency (MHz)\dotfill & 1400             & 1520        & 1374       \\
Number of beams\dotfill        & 1                & 1           & 13         \\
Integration time (min)\dotfill & 10               & 2.5         & 35         \\
Sample interval (ms)\dotfill   & 2.0              & 1.2         & 0.25       \\
Bandwidth (MHz)\dotfill        & $2\times8\times5$& $2\times64\times5$
                                 & $2\times 96\times3$                       \\
$S_{\rm sys}$ (Jy)\dotfill     & 60               & 70          & 36         \\
$S_{\rm min}$ (mJy)\dotfill    & 1.2              & 1.0         & 0.15       \\
Pulsars found/detected\dotfill & 40/61            & 46/100      & 513+/703+  \\
Reference\dotfill              & Clifton et al. & Johnston et al. & this work\\
\hline
\hline
\end{tabular}
\end{table}
Triggered by the above motivations, we are undertaking a new survey 
for pulsars along the Galactic plane at
1.4 Ghz, using the 13-element multibeam receiver recently installed
on the 64-m Parkes radiotelescope.  In this paper we present the experiment 
configuration, the survey plan
and the preliminary results of about 50\% of the survey.

\section{The Multibeam Survey}

Each beam of the multibeam receiver system at Parkes is approximately
0.23$^{o}$ wide and the beams centres are spaced 2 beamwidth apart 
(see Fig. 1). The survey pointings are interleaved to give complete sky
coverage on a hexagonal grid containing a total of 2670 pointings of 13
beams each. The parameters of the present experiment and those of two 
previous high frequency surveys
of the Galactic plane are summarized in Table 1.  Thanks to the long
integration time adopted (35-min) and the high sensitivity of the new receiver
system, the present survey has a sensitivity 7 times better than 
previous surveys.  Fig 2. shows the theoretical sensitivity as a function of the pulsar
period and dispersion measure.

\begin{figure}
\centerline{\psfig{file=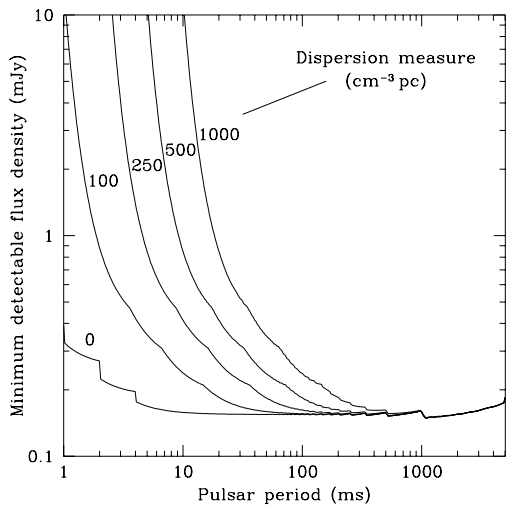, width=7cm}}
\caption[]{Theoretical sensitivity of the Parkes multibeam pulsar
survey as a function of period and DM for a pulsar on the centre beam.}
\end{figure}

\section{Results and discussion}

So far we have observed about 1600 pointings, 90\% of which are analysed,
corresponding
to about 50\% of the total survey region.  The data reduction system 
is similar to that used in the Parkes low frequency survey 
(Manchester et al, 1996), and is carried out on a network of workstations.
Because of the relatively long integrations adopted, we complement
the standard search analysis with ``acceleration search'' to take into account
possible binary motions.  To date we have discovered 513 new pulsars, and
have detected 190 known pulsars.  Accounting for the fact that so far
we searched the regions closest to the Galactic plane, we believe that
the number of new discoveries for the entire survey should be somewhat 
over 800.

Timing observations of the newly discovered pulsars are carried out at
Jodrell Bank and Parkes.  Observations are made at intervals of 4 -- 8
weeks, or more closely spaced when pulse-counting statistics need 
to be resolved.  Full timing solutions have been obtained so far for 80
pulsars.  At least eight of the new discoveries are young
pulsars ($\tau_{c}$ $<$ 10$^{5}$ years). Two radio pulsars with the 
highest known surface magnetic
field have been discovered (Kaspi et al, these proceedings).  One of
these objects, PSR J1119-6127 is very young, with as characteristic age,
$\tau_{c}$ = 1600 years.  For this pulsar we also measured a braking
index $n$=3.0$\pm$0.1.

So far, eight of the newly discovered pulsars proved to be members 
of binary systems, including a pulsar (PSR J1811-1736) in a highly eccentric 
binary system (Lyne et al 1999) and a pulsar (PSR J1740-3052) with a very
massive companion ($\simeq$ 11 M$_{\odot}$).  
The basic parameters of the binary pulsars are shown 
in Table 2.  

\setcounter{table}{1}

\begin{table}
\caption{New binary pulsars}
\begin{tabular}{lcccccc} \hline
PSR J & $P$ & $\tau_c$ & Distance & $P_b$ & Ecc. & Min. $M_c$ \\
      & (ms)& ($10^6$ y) & (kpc) & (d) &  & ($M_\odot$) \\ \hline
J1232$-$6501 &   !88.28 & 1400 &   10.0 &    !!1.863 &   0.00 &    !0.15 \\
J1904+04     &   !71.09 & --  &   !4.0 &    !15.750 &   0.04 &    !0.23 \\ 
J1810$-$2005 &   !32.82 & 4000  &   !4.0 &    !15.012 &   0.00 &    !0.29 \\
J1453$-$58   &   !45.25 & --  &   !3.3 &    !12.422 &   0.00 &    !0.88 \\
J1435$-$60   &   !!9.35 & --  &   !3.2 &    !!1.355 &   0.00 &    !0.90 \\ \\
J1811$-$1736 &   104.18 & !950  &   !5.9 &    !18.779 &   0.83 &    !0.87 \\
J1141$-$6545 &   393.90 & !!!1.45  &   !3.2 &    !!0.198 &   0.17 &    !1.01 \\ \\
J1740$-$3052 &   570.31 & !!!0.36  &   10.8 &    231.039 &   0.58 &    11.07 \\ \hline
\end{tabular}	 
\end{table}

\end{document}